\documentclass[aps,prl,twocolumn,amsmath,amssymb,nofootinbib,superscriptaddress]{revtex4-1}
\usepackage{times}
\usepackage[pdftex]{graphicx}
\usepackage{dcolumn}
\usepackage{bm}
\usepackage{amsmath}
\usepackage{indentfirst}
\usepackage{float}
\usepackage{ulem}
\usepackage[colorlinks]{hyperref}
\usepackage[dvipsnames]{xcolor}

\usepackage{algorithm}
\usepackage{algorithmic}

\definecolor{darkgreen}{rgb}{0.20,1,0.0}

\date{\today}

\begin{document}

\title{Ground state search by local and sequential updates of neural network quantum states }

\author{Wenxuan Zhang}
\affiliation{Science, Mathematics and Technology Cluster, Singapore University of Technology and Design, 8 Somapah Road, 487372 Singapore} 

\author{Xiansong Xu}
\affiliation{Science, Mathematics and Technology Cluster, Singapore University of Technology and Design, 8 Somapah Road, 487372 Singapore} 

\author{Zheyu Wu}
\affiliation{Science, Mathematics and Technology Cluster, Singapore University of Technology and Design, 8 Somapah Road, 487372 Singapore} 

\author{Vinitha Balachandran}
\affiliation{Science, Mathematics and Technology Cluster, Singapore University of Technology and Design, 8 Somapah Road, 487372 Singapore} 

\author{Dario Poletti} 
\email{dario\_poletti@sutd.edu.sg}
\affiliation{Science, Mathematics and Technology Cluster, Singapore University of Technology and Design, 8 Somapah Road, 487372 Singapore} 
\affiliation{Engineering Product Development Pillar, Singapore University of Technology and Design, 8 Somapah Road, 487372 Singapore} 
\affiliation{The Abdus Salam International Centre for Theoretical Physics, Strada Costiera 11, 34151 Trieste, Italy} 
\affiliation{Centre for Quantum Technologies, National University of Singapore 117543, Singapore}

\begin{abstract}  
Neural network quantum states are a promising tool to analyze complex quantum systems given their representative power. It can however be difficult to optimize efficiently and effectively the parameters of this type of ansatz. Here we propose a local optimization procedure which, when integrated with stochastic reconfiguration, outperforms previously used global optimization approaches. Specifically, we analyze both the ground state energy and the correlations for the non-integrable tilted Ising model with restricted Boltzmann machines. We find that sequential local updates can lead to faster convergence to states which have energy and correlations closer to those of the ground state, depending on the size of the portion of the neural network which is locally updated. To show the  generality of the approach we apply it to both 1D and 2D non-integrable spin systems.    
\end{abstract}

\date{\today}

\maketitle

\begin{figure}
\includegraphics[width=\columnwidth]{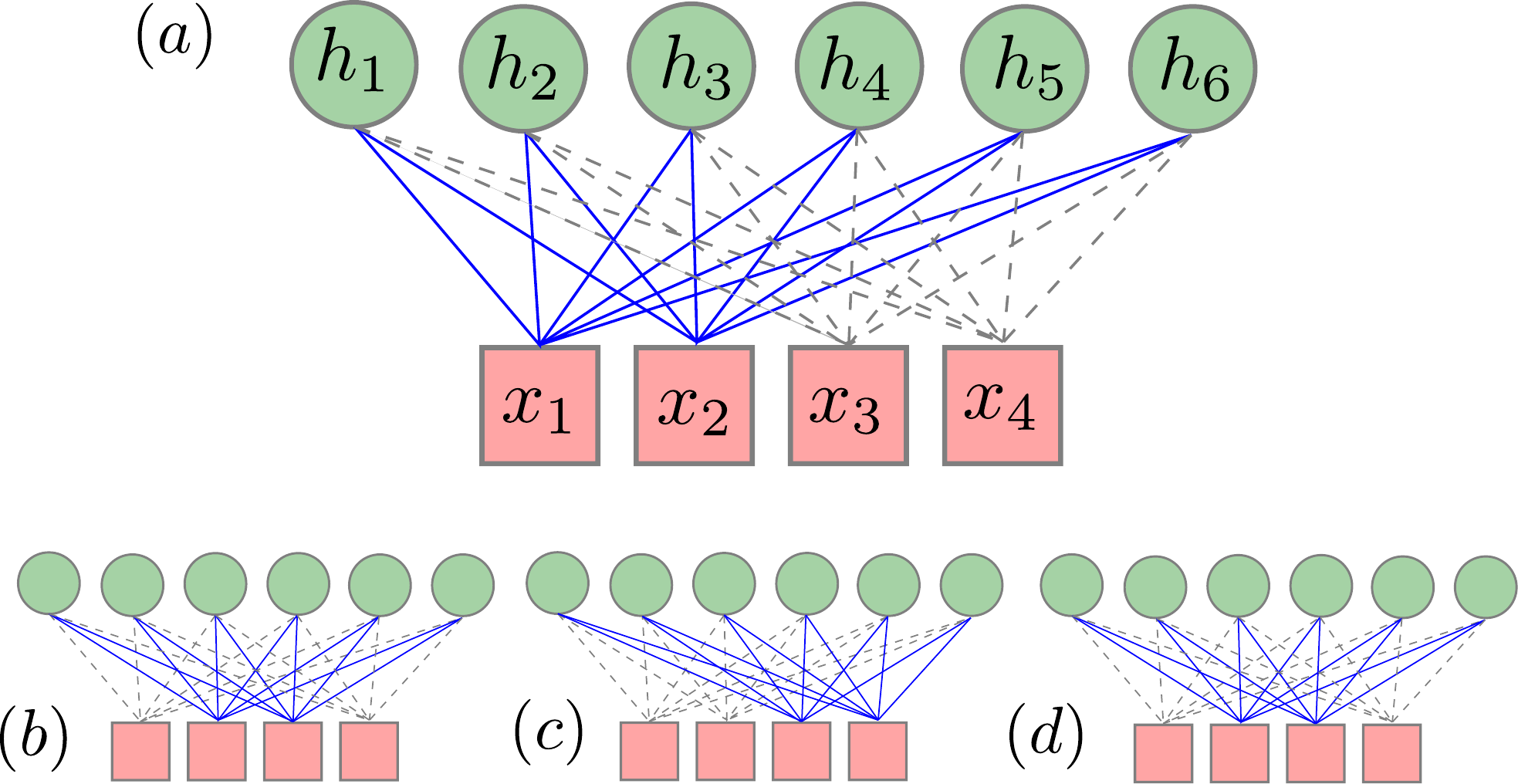} 
\caption{(a) Depiction of a restricted Boltzmann machine. The red squares represent physical spins while the green circles represent hidden nodes. The lines represent the links of the neural network. (a) to (d) A sweep of the sequential local optimization algorithm for a system with $4$ visible nodes and a block size of $2$ visible nodes. At each step, the links which are optimized, represented by the blue continuous lines, are updated.} \label{fig:fig1} 
\end{figure}

{\it Introduction}: Recent years have seen a significant increase in the use of machine learning and neural networks in the physical sciences including statistical physics, particle physics, cosmology, many-body quantum physics, quantum computing, quantum processes and quantum chemistry \cite{CarleoZdeborova2019, MelkoCirac2019, CarrasquillaTorlai2021, BandyopadhyayZhao2018, rem2019identifying, yang2020applications, BanchiSeverini2018, LuchnikovFilippov2019, biamonte2017quantum, beer2020training, carrasquilla2017machine, radovic2018machine, perraudin2019deepsphere, GuoPoletti2020, BaireyArad2020, FlurinSiddiqi2020, LinLan2021}. In particular, Carleo and Troyer first demonstrated how a neural network quantum state can be applied together with variational quantum Monte Carlo \cite{carleo2017solving}, where they computed the ground state and the dynamics of a system with a restricted Boltzmann machine (RBM) \cite{smolensky1986information}, see Fig.~\ref{fig:fig1}(a) for a depiction of an RBM.  Together with RBMs \cite{MelkoCirac2019, ZenBressan2020, ZenBressan2020b, NomuraImada2017,VieijraVerstraete2020, GolubevaMelko2022, PilatiPieri2020, LuDuan2019, NomuraNori2021, VicentiniCiuti2019, YoshiokaHamazaki2019, NagySavona2019, HartmannCarleo2019, Nomura2021, ParkKastoriano2020, ParkKastoriano2021, ChooNeupert2018}, other network structures such as a feed-forward \cite{cai2018approximating, ChooNeupert2018, LuoClark2019, SharirShashua2020, KesslerKuhne2021}, recurrent \cite{HibatAllahCarrasquilla2020, Roth2020} and convolutional neural networks \cite{choo2019two, schmitt2020quantum, GutierrezMendl2022, LiPu2020, IrikuraHiroki2020, VieijraNys2021, LiangHe2018, LiuWang2021, SaitoKato2021, RothMacDonald2021, FuJi2021, LiangWei2022} have also been intensively studied. The significant interest in the field is attributed to the fact that these networks could offer possibility to fight against the curse of dimensionality in many-body quantum systems. For example, RBMs have been shown to be able to represent volume law states \cite{deng2017quantum} and to correspond to tensor networks with exponentially large bond dimensions \cite{ChenXiang2018, ColluraMontangero2021}. 

Different strategies have been studied in order to improve the accuracy and speed of convergence of neural network quantum states algorithms, such as transfer learning \cite{ZenBressan2020}, pruning \cite{GolubevaMelko2022}, importance sampling \cite{LiPu2020}, and massive parallelization \cite{LiangWei2022}. Between these different techniques, it has been consistently shown that stochastic reconfiguration \cite{sorella2007weak}, effectively a Hessian based optimization routine motivated by imaginary time evolution, can provide significantly better performance. However, stochastic reconfiguration requires solving a linear problem whose size increases quadratically with the system size, thus making this process challenging for large systems. It is thus important to improve its scalability so that we can tackle large and high dimensional systems. 

It must be mentioned that, long before neural network quantum states were proposed, a large number of highly performing methods have been devised for many-body quantum physics, each with its own strengths. Two very successful examples are (exact or variational) quantum Monte Carlo \cite{MolmerDalibard1993, DumRitsch1992, DumGardiner1992, PlenioKnight1998, mcmillan1965ground}, and tensor network methods \cite{verstraete2008matrix, orus2014practical, SilviMontangero2019}. The current implementation of ground state search with neural network quantum states follows closely variational quantum Monte Carlo, with the important point that the variational wave function is described by a neural network. Tensor network methods, which also fall into the category of variational methods, allow describing accurately ground states, and they have emerged as the best method in 1D systems with local Hamiltonians. Typically, the ground state search with tensor network methods, e.g., with matrix product states, is implemented in an iterative manner in which one only locally optimizes the tensor network parameters and then ``sweeps'' through the system thus approaching the ground state after a number of minimization steps which scales linearly with the system size. In particular, it has been shown that, for tensor networks, local optimizations generally result in better performance than global optimizations \cite{LiuDeng2021}. While interesting developments in the field of tensor networks are continuously appearing \cite{VanderstraetenVerstraete2022, LiaoXiang2019, FerrariMontangero2022}, important limitations still exist, for example for dimensions larger than one and for long range systems.

In this work we bring together the variational quantum Monte Carlo inspired neural network states algorithm with stochastic reconfiguration, and the local minimization procedure of tensor networks. 
As depicted in Fig.~\ref{fig:fig1}, we choose a portion of the system to be minimized, e.g. two sites in the figure, and iteratively update only the block of parameters of the neural network which are directly coupled to these sites. We will then optimize another group of sites which, importantly, has a sizeable overlap with the previous ones. We perform the local optimizations sequentially until we reach the end of the system and then we move backward, see an example in Fig.~\ref{fig:fig1}(a) to (d). In analogy to tensor network algorithms, we refer to this sequential optimization procedure as a sweep, while for the algorithm we will use the expression {\it sequential local optimization} (SLO).  
To evaluate the performance of this algorithm, we consider a non-integrable Ising model with transverse field (equivalent to interacting spinless fermions), and we apply the algorithm in the three different phases (ferromagnetic, paramagnetic and anti-ferromagnetic). We then evaluate how fast and how accurately we can describe both the energy and the correlation functions of the ground state.  

We find that SLO improves dramatically the performance compared to previously commonly used global optimization algorithms both in terms of speed of convergence and accuracy of the ground state found. We also apply this method to large 1D systems and 2D setups, showing the generality of the speed-up obtained from SLO.


{\it Restricted Boltzmann Machine}: For a concrete example of neural networks, in this work we use RBMs. They are composed of a single visible layer of $L$ nodes (one for each spin in the system) and a single hidden layer of $M=\alpha L$ nodes where $\alpha$ is a natural number hyperparameter. The input physical configuration and bias of visible layer are denoted with $\pmb{x}=\left[\sigma_1, \sigma_2, ..., \sigma_L\right]^{\rm T}$ and $\pmb{a}$, respectively, and the hidden nodes are characterized by the auxiliary spin configuration $\pmb{h} = \left[ h_1, h_2, ... , h_M \right]^{\rm T}$. Here visible and hidden nodes only take the values $\pm 1$. The visible nodes are connected to the hidden ones via the weights matrix $\pmb{W} =(W_{ji}) \in \mathbb{C}^{M \times L}$. Thus, given the configurations $\pmb{x}$ and $\pmb{h}$, one can readily compute a complex amplitude \footnote{Typically one would also add a bias for the hidden nodes, but for the systems we consider it does not significantly improve the performance.}
\begin{align}
f(\pmb{x,h};\pmb{a,W}) = \exp(\pmb{a}^{\rm T} \pmb{x} +  \pmb{h}^{\rm T} \pmb{W x }). \label{eq:RBM_f}  
\end{align}
Since there is no connection for nodes within a layer, we can readily trace out the auxiliary configurations of the hidden layer to obtain the ansatz for the wave function explicitly
\begin{align}\label{wavefunction_1}
	\psi(\pmb{x} ; \pmb{a,W}) & = \sum_{\{h_i\}}f(\pmb{x,h;a,W})  \nonumber\\ 
	& = \exp(\pmb{a}^{\rm T} \pmb{x} )\prod_j^M 2\cosh( {\sum_i W_{ji}x_i}).  
\end{align}
The above analytical form for the variational ansatz allows us to derive an analytical expression for its derivatives with respect to all the parameters $\pmb{\theta} \equiv \{\pmb{a}, \pmb{W}\}$.
For this to be done, however, one needs to evaluate the network for all the possible configurations, which is not scalable as they grow in number exponentially with the system size. One does revert to a probabilistic approach, only considering the most likely configurations with their corresponding frequency which we evaluate with a Metropolis-Hasting algorithm \cite{metropolis1953equation}.

{\it Optimization with stochastic reconfiguration}: The update of the network parameters $\pmb{\theta}(l)$ at the $l\text{-}$th iteration is given, following stochastic reconfiguration, by   
\begin{equation}
	\pmb{\theta}{(l+1)} = \pmb{\theta}(l) - \gamma(l)  \pmb{S}^{-1}(l) \pmb{F}(l). \label{eq:SR} 
\end{equation}
The above $\pmb{F}$ is the gradient $F_k = \langle E_{loc} O_k^*\rangle-\langle E_{loc}\rangle \langle O_k^*\rangle$ where $O_k = \partial\ln\psi(\pmb{x};\pmb{\theta})/\partial \theta_k$ and $E_{loc}(\pmb{x}) = \sum_{\pmb{x'}}H_{\pmb{x}\pmb{x'}}\frac{\psi(\pmb{x'};\pmb{\theta})}{\psi(\pmb{x};\pmb{\theta})}$, while $\gamma(l)$ is the learning rate at $l\text{-}$th iteration. What sets apart stochastic reconfiguration update in Eq.(\ref{eq:SR}) from a stochastic gradient descent is the presence of the covariance matrix $\pmb{S}$, given by $S_{km} = \langle O^*_kO_m\rangle - \langle O^*_k\rangle\langle O_m\rangle$. 
For an RBM, the size of $\pmb{S}$ is $(L+\alpha L^2)\times (L+\alpha L^2)$. Hence, a much large amount of time is required to compute $\pmb{S}^{-1}$ for larger systems as it would typically scale as $L^{4.6}$ \cite{coppersmith1987matrix}. Hence, for large system sizes one can lose the benefits of using stochastic reconfiguration.   

{\it Sequential Local Optimization}: 
To circumvent this, we propose a sequential local optimization (SLO) approach based on considering, each time, only a portion of the parameters of the network, i.e. $\pmb{\theta}^{p,s}$ where $p$ indicates the position of its leftmost (and lowermost in 2D) visible node, and $s$ represents the size of the group of visible nodes considered. For instance, in Fig.~\ref{fig:fig1}(a) we have $s=2$, while for a 2D system we could have $s=2\times 2$. It results that the corresponding covariance matrix $\pmb{S}^{p,s}$ has dimension $s(\alpha L+ 1) \times s(\alpha L+1)$ which implies a significant speed up in computing the inverse.
The sequential optimization is performed in a similar manner to DMRG sweeps \cite{Schollwock2005}, whereby one optimizes different blocks $\pmb{\theta}^{p,s}$ from left to right and back to left (for 1D), ensuring that consecutive $\pmb{\theta}^{p,s}$ which are optimized partially overlap. An example of the sequential update for a complete sweep for a system with 4 visible nodes and $s=2$ with an overlap of one node between consecutive parameters blocks is depicted in Fig.~\ref{fig:fig1}(a) to (d).          
Naturally, the smaller the blocks considered, the larger is the number of minimization steps. However, in each sweep this only scales linearly with $L$ and thus there is a possibility for SLO to perform better with smaller block sizes. 
In the following we study the performance of local optimization versus the size of the system, the size of the blocks and the Hamiltonian parameters which can bring the ground state in different phases of matter. In the following, unless explicitly stated otherwise, we choose the following hyperparameters, which allow us to approach the ground state for all system and block sizes $s$ considered: $\alpha=5$, initial and final learning rate respectively $\gamma_0=0.1$ and $\gamma_f=0.0125$, with a decrease of a factor $0.5$ every two sweeps; the number of samples in each iteration is $N_s=10^4$ and the number of thermalization steps to generate the first set of configurations is $N_{thermal} = 100$.  
Since the optimization procedure is stochastic, the results may vary if it is run more than once. This is particularly important when considering systems in which the ansatz can be trapped in local minima (e.g. when evaluating the ground state in the ferromagnetic phase or antiferromagnetic phase ). Hence, for each ground state to be computed, we ran the optimization algorithm $20$ times and recorded the best result. 
Our algorithm has been written in a Julia code which runs all the samplings and optimization on GPUs, communicating with CPUs only when outputting results. The GPU used is an NVIDIA A100 40G.

{\it Model}: To evaluate the SLO in different phases, we consider both a 1D and 2D spin$-1/2$ tilted Ising models (TIM) 
\begin{align}
H = \sum_{\langle i,j\rangle} J\sigma^z_i\sigma^z_{j} - \sum_{i} (h_z \sigma^z_{i}  + h_x \sigma^x_{i}).   
\end{align}   
where $h_x$ and $h_z$ are local fields, $J$ represents the interaction between the spins and sum over $\langle i, j\rangle$ means on neighbouring sites. The $\sigma^a_i$, with $a=x,y,z$ are the spin$-1/2$ operators acting on spin $i$ which can be represented by Pauli matrices.  
When the local fields are small enough, the TIM is either in the ferromagnetic (FM) phase, when the value of $J$ is negative, or in the antiferromagnetic (AFM) phase, when $J$ is positive while, when local fields are relatively strong, a paramagnetic (PM) phase emerges. Note that we always consider nonzero $h_x$ and $h_z$ so that the system is non-integrable. 
In the following we work in units for which $|J|=1$. 
We will use open-boundary conditions as they require a larger number of hidden nodes than the periodic boundary conditions for which one can reduce the number of hidden nodes from $\alpha L$ to $\alpha$  \cite{carleo2017solving}.

{\it Analysis}: 
We study how different choices of SLO converge towards the ground state both as a function of sweeps and of running time $t_r$ in seconds. 
This is because when we use smaller blocks we also need to do more minimization per sweep. 
To evaluate the quality of the ground state found, we study both the energy, given by $E = \langle\psi|H|\psi\rangle$, and the correlators in the system $C_d^F$ and $C_d^A$, respectively the ferro and antiferro-magnetic correlators, given by     
\begin{align}
	C_d^F &= \frac{1}{d}\sum_{l=2}^{d+1} \langle \sigma^z_1\sigma^z_l \rangle \label{eq:ferro}\\ 
	C_d^A &= \frac{1}{d}\sum_{l=2}^{d+1} (-1)^{l-1} \langle \sigma^z_1\sigma^z_l \rangle.  \label{eq:antiferro} 
\end{align} 
Here $\langle\sigma^z_1\sigma^z_k \rangle$ indicates the expectation value of the spin-spin correlation in the $z$ direction between the first and the $k$-th spins, and $d$ is the distance considered.

\begin{figure}
\includegraphics[width=\columnwidth]{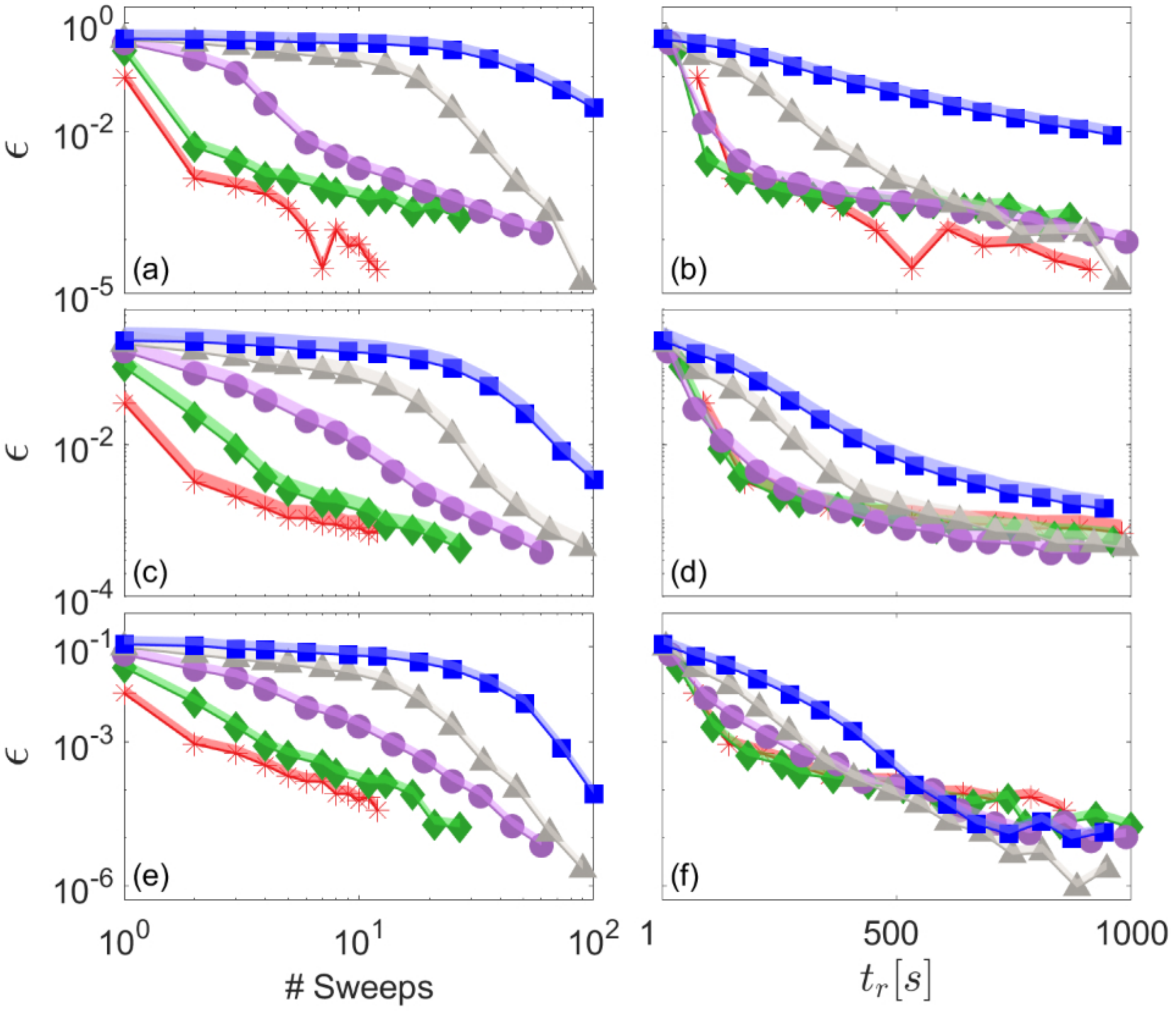}
\caption{Convergence performance of ground state energy for 1D, 32 spins TIM model with open boundary conditions. Common parameters are $J = 1$ and $h_z = 0.5$. The phase of the system is varied with different field strength $h_x$: panel (a) and (b): antiferromagnetic phase with $h_x = 0.5$; panel (c) and (d): phase transition point with $h_x = 0.95$; panel (e) and (f): paramagnetic phase with $h_x = 1.5$. The symbol: \textcolor{red}{$\ast$} represents 2-sites block; \textcolor{ForestGreen}{$\blacklozenge$} represents 4-sites block; \textcolor{Purple}{$\bullet$} represents 8-sites block; \textcolor{gray}{$\blacktriangle$} represents 16-sites block; \textcolor{blue}{$\blacksquare$} is for all 32-sites; for the different block size: $s = \{2, 4, 8, 16\}$, the respective number of iterations in each sweep is $n_s = \{60, 28, 12, 4\}$. The left column panels represent the accuracy $\epsilon$ versus number of sweeps, while the right column panels indicate the accuracy versus computational time. Finally, the light-colored shadow for each curve represents the region of the lowest 50$\%$ of values from 20 runs. } \label{fig:fig2} \end{figure}

{\it Results}: In Fig.~\ref{fig:fig2} we consider the relative error of ground energy $\epsilon = |E - E_{gs}|/|E_{gs}|$ where $E_{gs}$ has been obtained, using matrix product states calculations and studying the convergence of the ground state energy for bond dimensions from 20 up to 800, with an absolute accuracy of $10^{-8}$. In panels (a,c,e) we plot $\epsilon$ versus the number of optimization sweeps, while in (b,d,f) we plot it versus the running time $t_r$.  In panels (a,b) we consider Hamiltonian parameters such that the ground state is in the antiferromagnetic phase, in panels (e,f) for the ground state in the paramagnetic phase and (c,d) are close to the transition point between the paramagnetic and antiferromagnetic phases. In each panel we consider different block sizes. With blue squares we represent the case of a single block which is as large as the system (i.e. the usual stochastic reconfiguration procedure), and we also consider blocks of size $16$, gray triangles, $8$, purple circles, $4$, green diamonds and $2$, red stars.       
We can observe that the performance of SLO is generally better than the strategy of updating all parameters both as a function of the number of sweeps, and as a function of running time. This is particularly evident in the antiferromagnetic phase and at the transition, where it is much harder for the RBM to approach the ground state compared to the paramagnetic phase.

\begin{figure}
\includegraphics[width=\columnwidth]{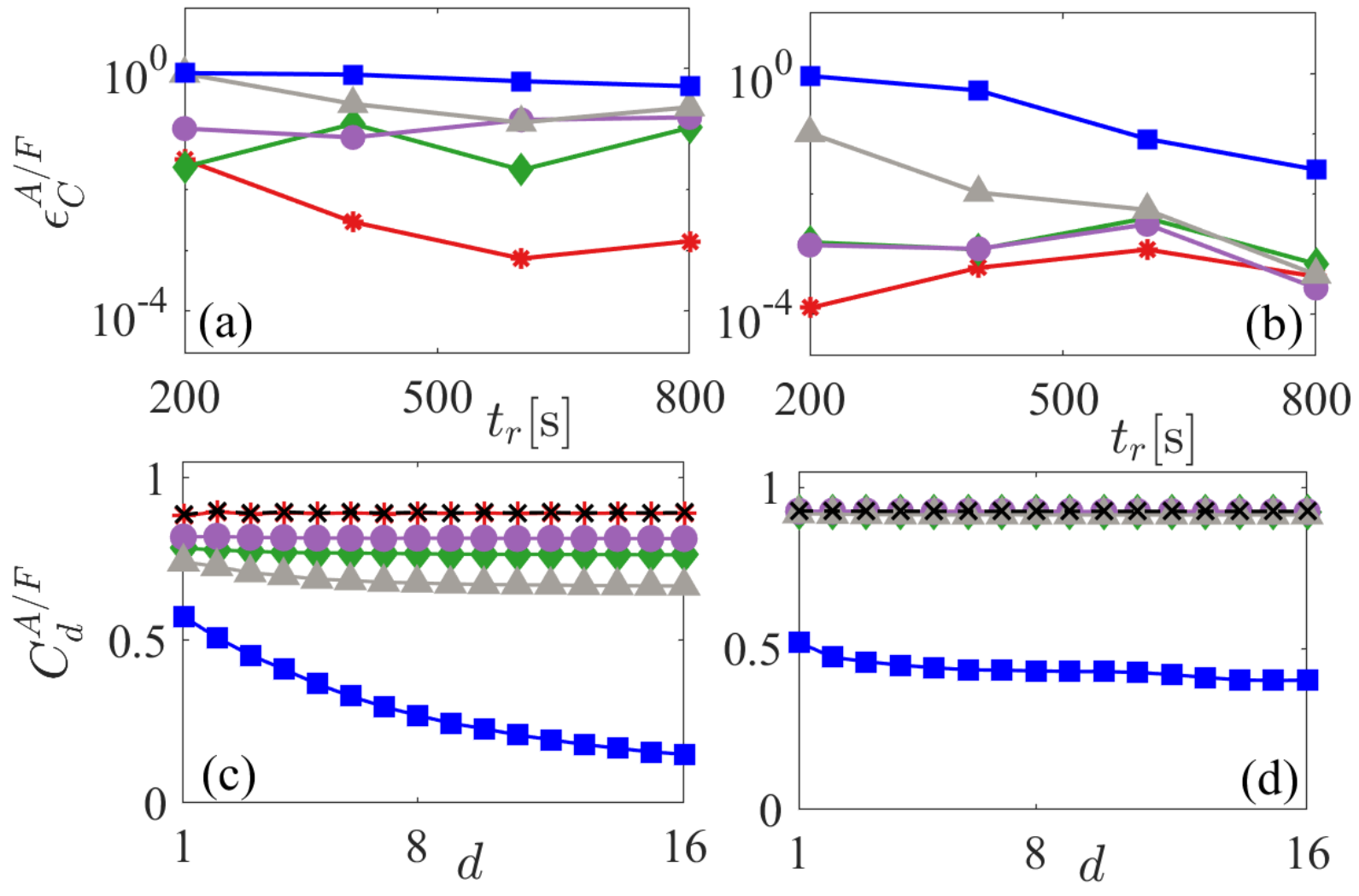} 
\caption{Convergence performance of correlators for the ground state of two phases: antiferromagnetic phase for panel (a) and (c) where $J = 1, h_x = h_z = 0.5$ and ferromagnetic phase for panel (b) and (d) where $J = -1, h_x = h_z = 0.5$. Panel (a) and (b) show the accuracy of correlators with $d =31$ versus time. Panels (c) and (d) show the efficiency of correlators when the simulation time is 400s. 
The symbol \textcolor{red}{$\ast$} represents 2-site block, \textcolor{ForestGreen}{$\blacklozenge$} 4-site block, \textcolor{Purple}{$\bullet$} 8-site block, \textcolor{gray}{$\blacktriangle$} 16-site block and \textcolor{blue}{$\blacksquare$} represents updating all parameters. The black continuous line in panels (c,d) represents the value from the MPS computation.}   \label{fig:fig3} 
\end{figure}

In Fig.~\ref{fig:fig3} we focus on the ability of reproducing the correct ground state correlations. We thus consider parameters that give an antiferromagnetic or a ferromagnetic ground state respectively in Fig.~\ref{fig:fig2}. In Fig.~\ref{fig:fig3}(a,b) we show the error as a function of time of $C^A_{L-1}$ and $C^F_{L-1}$, respectively, when compared to the results from matrix product states. More precisely we plot $\epsilon^{A/F}_C = |C^{A/F}_{L-1} - C^{A/F}_{L-1, gs}|$ where $C^{A/F}_{L-1, gs}$ is computed using our matrix product states algorithm. 
Given a fix amount of time of simulation $t_r = 400s $, we plot $C^{A/F}_{d}$, versus $d$ for antiferro (c) and ferro (d) ground states. In all the panels  the SLO algorithm with $s=2$ consistently approaches the matrix product states results better than larger block sizes.      
We clarify that, in all these panels, the value of the correlations used corresponds to the run which has the lowest energy out of the 20 runs.

\begin{figure}
\includegraphics[width=\columnwidth]{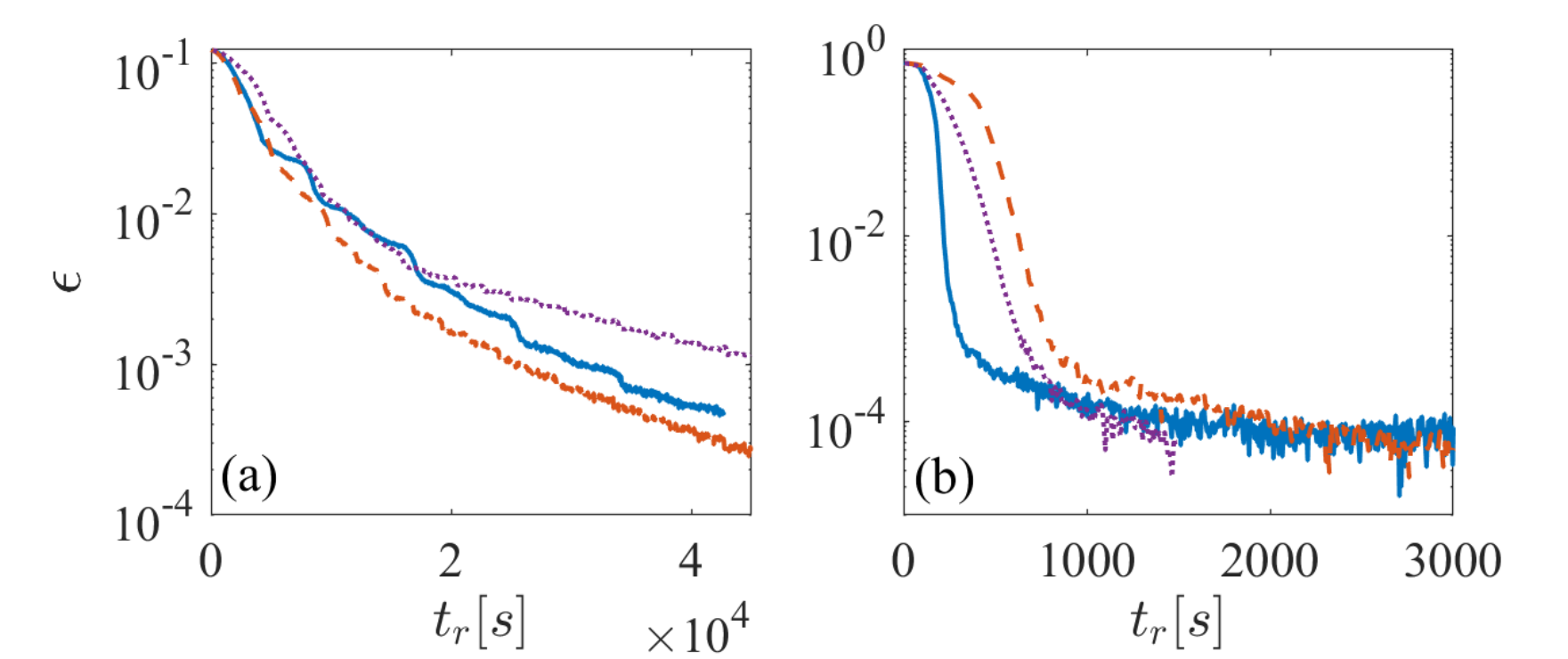} 
\caption{Relative error $\epsilon$ versus running time for a TIM with (a) 128 spins in 1D with $h_x = 1.5$, $h_z = 0.5$ and (b) $8\times 8$ spins in 2D  with $h_x = 0.5$, $h_z = 0.5$. (a) Blue continuous line corresponds to $s=2$, red dashed line to $s=4$ and purple dotted line to $s=16$. (b) Blue continuous line corresponds to $s=2\times 2$, red dashed line to $s=3\times 3$ and purple dotted line to $s=4\times 4$.  }   \label{fig:fig4} 
\end{figure}

Until now we have considered relatively small 1D systems. This was so that we could obtain in reasonable times the ground state even if we were taking blocks of the size of the whole system, as we wanted to compare the performance of different block sizes. Now we consider larger systems such that we can investigate the effectiveness of SLO also for larger 1D and 2D systems. In Fig.~\ref{fig:fig4}(a) we show the relative error for the 1D TIM for $128$ spins with $10^5$ samples. Each line corresponds to different block sizes, $s=2$ for blue continuous line, $4$ for the red dashed line and $16$ for the purple dotted line \cite{learningrate128}. The reference ground state energy value has been computed with a matrix product states algorithm. We observe that, smaller block sizes tend to reach better precision faster than block size 16, which is the largest we have considered here. We then considered an $8\times 8$ 2D TIM with open boundary conditions. Here, to obtain a ground state energy reference value we use a neural network with higher representative power, i.e. $\alpha = 7$, more accurate values for the gradients and the correlation matrices by using $40000$ samples, and we checked the convergence with different block sizes $s$. This analysis gives us, for $h_x=h_z=0.5$, a reference ground state energy of $-114.44$ to five significant digits. We then run simulations with $\alpha =5$ and $10000$ samples using block sizes $s=2\times 2$ (blue continuous line), $3\times 3$ (red dashed line) and $4\times 4$ (purple dotted line), as shown in Fig.~\ref{fig:fig4}(b) \cite{learningrate64}. We can see that SLO allows to reach the ground state with good accuracy and smaller blocks can allow a faster descent of the energy value towards the ground state's one.    

{\it Conclusions}: We have introduced an optimization algorithm for neural network quantum states which is based on local optimization and which, in general, allows us to obtain more accurate ground states in shorter times. 
We have tested this by computing both energy and correlations in different phases of matter and dimensionalities of the system.

We now highlight a few possible outlooks.  
The current approach can be readily translated to other neural networks where one can optimize blocks of the network which are coupled to a portion of the system.    
For even larger systems than the ones we considered, the number of hidden nodes may be too large. It is however possible to use a segmentation of the hidden nodes and perform sweeps both on the visible and hidden nodes.  
Another aspect that should be investigated in the future is the role of the locality of the Hamiltonian used, and in particular the effectiveness of the size $s$ of the SLO versus the range of the couplings in the Hamiltonian. 
Study of frustrated systems which require neural networks with better variational sign structures \cite{choo2019two, NomuraImada2017, NomuraImada2021, FerrariCarrasquilla2019, SzaboCastelnovo2020, LiangHe2021, ChenNeupert2022}; Linear methods for the optimization of RBMs which require less epoch, but each epoch is more demanding, have been proposed \cite{FrankKastoriano2021}. Depending on the systems studied and the difficulty of sampling, these methods can be advantageous;  
We should also mention that it is possible to use stochastic reconfiguration and compute the gradient on the fly iteratively \cite{VicentiniCarleo2021}, but this approach can be costly when dealing with ill-conditioned matrices, and it is still a global approach. Detailed comparisons with this method should be pursued in the future;  
Finally more work can be done in finding the most efficient ways to do partial optimizations of the neural networks, and on how to tune its hyperparameters, such as the learning rate.

\begin{acknowledgments} 
We are grateful to Bo Xing and Marcello Dalmonte for fruitful discussions. D.P. acknowledges support from the Ministry of Education Singapore, under the grant MOE-T2EP50120-0019. The computational work for this article was partially performed on the National Supercomputing Centre, Singapore \cite{NSCC}.  
\end{acknowledgments}

\normalem
\bibliographystyle{apsrev4-1}
\bibliography{Bibliography.bib}

\begin{thebibliography}{82}
\bibitem [{\citenamefont {Here}()}]{learningrate128}%
  \BibitemOpen
  \bibfield  {author} {\bibinfo {author} {\bibnamefont {Here}},\ }\href@noop {}
  {}\bibinfo {howpublished} {the learning rate goes from $0.08$ to $0.02$. The
  learning rate is halved every 70, 30, and 30 iterations respectively for
  $s=2$, $4$ and $16$.}\BibitemShut {Stop}%
\bibitem [{lea()}]{learningrate64}%
  \BibitemOpen
  \href@noop {} {}\bibinfo {howpublished} {Here, the learning rate goes from
  $0.08$ to $0.02$. The learning rate is halved every 50, 8, and 8 iterations
  respectively for $s=2\times 2$, $3 \times 3$ and $4 \times 4$.}\BibitemShut
  {Stop}%
\end{thebibliography}

\end{document}